\documentclass[12pt]{article}
\usepackage{cite}
\usepackage{graphicx}
\usepackage{amssymb,amsfonts}
\usepackage{hangcaption}

\renewcommand{\theequation}
{\arabic{section}.\arabic{equation}}

\makeatletter
\def\eqnarray{ \stepcounter{equation} \let\@currentlabel=\theequation
 \global\@eqnswtrue
 \global\@eqcnt\z@
 \tabskip\@centering
 \let\\=\@eqncr
 $$\halign to \displaywidth\bgroup\@eqnsel\hskip\@centering
 $\displaystyle\tabskip\z@{##}$&\global\@eqcnt\@ne
 \hfil$\displaystyle{{}##{}}$\hfil
 &\global\@eqcnt\tw@$\displaystyle\tabskip\z@{##}$\hfil
 \tabskip\@centering&\llap{##}\tabskip\z@\cr}
\makeatother

\makeatletter
\def\@arrayacol{\edef\@preamble{\@preamble \hskip .5\arraycolsep}}
\def\array{\let\@acol\@arrayacol \let\@classz\@arrayclassz
\let\@classiv\@arrayclassiv \let\\\@arraycr\def\@halignto{}\@tabarray}
\makeatother

\makeatletter
\newcounter{subeqncnt}
\def\thesubeqncnt{\alph{subeqncnt}}
\def\subequations{\begingroup%
   \stepcounter{equation}\edef\@tempa{\theequation}%
   \let\c@equation\c@subeqncnt\c@subeqncnt\z@
   \edef\theequation{\@tempa\noexpand\thesubeqncnt}}

\makeatother

\newcommand{\captionfonts}{\small}
\makeatletter 
\long\def\@makecaption#1#2{%
\vskip\abovecaptionskip
\sbox\@tempboxa{{\captionfonts #1: #2}}%
\ifdim \wd\@tempboxa >\hsize {\captionfonts #1: #2\par} \else
\hbox to\hsize{\hfil\box\@tempboxa\hfil}%
\fi \vskip\belowcaptionskip}
\makeatother 

\newcommand{\del}{\partial}

\newcommand{\dd}{{\rm d}}

\def\imo{i}

\begin{document}

\setlength{\baselineskip}{7mm}

\begin{flushright}
{\tt SHU-Pre2009-08} \\
{\tt arXiv:0905.2675[hep-th]} \\
August, 2009
\end{flushright}

\vspace{1cm}

\begin{center}
{\Large Shear viscosity and instability from third order Lovelock
gravity}

\vspace{1cm}

{\sc{Xian-Hui Ge}}$^*$, {\sc{Sang-Jin Sin}}$^{\dagger}$,
{\sc{Shao-Feng Wu}}$^*$\\
and {\sc{Guo-Hong Yang}}$^ *$

$*${\it{Department of Physics, Shanghai University},} \\
{\it{Shanghai 200444, China}} \\
{\sf{gexh@shu.edu.cn, sfwu@shu.edu.cn,ghyang@shu.edu.cn}}
\\
$\dagger$ {\it{Department of Physics,}} {\it{Hanyang University,}}
{\it{Seoul 133-791, Korea}} \\
{\sf{sjsin@hanyang.ac.kr}}
\end{center}

\vspace{1.5cm}

\begin{abstract}
We calculate the ratio of shear viscosity to entropy density for
charged black branes in third order Lovelock theory. For chargeless
black branes, the result turns out to be consistent with the
prediction made in $\rm arXiv:0808.3498[\rm hep-th] $. We find that,
the third order Lovelock gravity term does not contribute to
causality violation  unlike the Gauss-Bonnet term. The stability of
the black brane again requires the value of the Lovelock coupling
constant to be bounded by $1/4$ in the infinite dimensionality
limit.
\end{abstract}

\section{Introduction}
\setcounter{equation}{0} \setcounter{footnote}{0}

The AdS/CFT correspondence \cite{ads/cft,gkp,w} provides an
interesting theoretical framework for studying relativistic
hydrodynamics of strongly coupled gauge theories. The result of RHIC
experiment on the viscosity/entropy ratio turns out to be in favor
of the prediction of AdS/CFT ~\cite{pss0,kss,bl}. Some attempt has
been made to map the entire process of RHIC experiment in terms of
gravity dual \cite{ssz}. The way to include chemical potential in
the theory was figured out in~\cite{ksz,ht}.

It had been conjectured that the viscosity value of theories with
gravity dual may give a lower bound for the $\eta/s=\frac{1}{4 \pi}$
for all possible liquid \cite{kovtun}. However, in the presence of
higher-derivative gravity corrections, the viscosity bound and
causality are also violated as a consequence
\cite{kp,shenker,shenker1,neupane}. The ratio of shear viscosity to
entropy density are of particular interest in higher derivative
gravity duals because those higher derivative terms can be regarded
as generated from stringy corrections given the vastness of the
string landscape. In \cite{brustein1,brustein2,Iq}, the authors
computed $\eta/s$ for general gravity duals by determining the ratio
of two effective gravitational couplings. The $\eta/s$ in presence
of arbitrary $R^2$
 and $R^3$ terms in the bulk action were calculated in
 \cite{ban}.

The higher derivative terms may be a source of inconsistencies
because higher powers of curvature could give rise to fourth or even
sixth order differential equation for the metric, and in general
would introduce ghosts and violate unitarity. Zwiebach and Zumino
\cite{zw,zu} found that ghosts can be avoided if the higher
derivative terms only consist of the dimensional continuations of
the Euler densities, leading to second order field equations for the
metric. These theories are the so called Lovelock
gravity\cite{lovelock}. The zeroth order of Lovelock gravity
correspondences to the cosmological constant. The first order is the
Einstein equation and the second order correspondences to
Gauss-Bonnet theory. Higher derivative effects on $\eta/s$ in the
presence of a chemical potential have been
 discussed in \cite{gmsst,cai2,cai3,ges,Jliu,merys,fada,McInnes}. In this paper, we discuss shear viscosity in
third order Lovelock gravity.

Our motivation for this paper is based on the following facts:

1). Although people expect that $\eta/s$ might receive corrections
from third and higher order Lovelock terms, it was conjectured in
\cite{brustein1} that $\eta/s$ gets no corrections at all for higher
order Lovelock terms except the Gauss-Bonnet terms. In this paper,
we compute $\eta/s$ for third order Lovelock gravity directly by
using the standard  method developed in \cite{pss0,kss} and compare
our result with that of \cite{brustein1}.

2). In \cite{kp} and \cite{shenker}, the authors showed that if we
consider the Gauss-Bonnet correction to Einstein equation, the
viscosity bound is violated in the hydrodynamics regime. Moreover,
causality violation happens in the high frequency regime
($k^{\mu}\rightarrow \infty$), which implies that theories in that
regime are pathological\cite{shenker1}{\footnote{The causality issue
in Gauss-Bonnet gravity was further studied in \cite{Buchel}.}}. In
\cite{gmsst} and \cite{ges}, some of us considered medium effect and
the higher derivative correction simultaneously by adding charge and
Gauss-Bonnet terms and found that the viscosity bound as well as
causality violation is not changed by the charge. After adding the
third order Lovelock terms, the causality structure would be
different from that of the Gauss-Bonnet gravity.

3). It is worth to studying the stability of black branes (black
holes) in third order Lovelock gravity in that the stability can
 constrain the higher derivative coupling constants. The
instability of $D$-dimensional asymptotically flat
Einstein-Gauss-Bonnet and Lovelock black holes has been discussed by
several authors\cite{dotti,konoplya,taka,deh}. Their results show
that for  gravitational perturbations of Schwarzschild black holes
in $D\geq 5 $ Gauss-Bonnet gravity, instability occurs only for
$D=5$ and $D=6$ cases at large value of $\alpha'$ \cite{konoplya}.
In \cite{taka}, the authors showed that small black holes in
Lovelock gravity are unstable. In this paper, we extend our previous
computation to third order Lovelock gravity in $D$-dimensional
spacetime and show how stability constrains the Lovelock coupling
constant.

The plan of this paper is as follows. In section 2, we briefly
review the thermodynamic properties of Reissner-Nordstr\"om-AdS
black brane solution in third order Lovelock gravity. In section 3,
we compute the viscosity to entropy density ratio via Kubo formula
and its charge dependence. In section 4, the causality problem is
discussed. We study the stability issue of  Reissner-Nordstr\"om-AdS
black branes in third order Lovelock gravity in section 5.
Conclusions and discussions are presented in the last section.

\section{Reissner-Nordstr\"om-AdS black brane in third order Lovelock gravity}
\setcounter{equation}{0} \setcounter{footnote}{0}

We start by introducing the following action in $D$ dimensions which
includes Lovelock terms and $U(1)$ gauge field:
\begin{equation}
\label{lovelockaction} I=\frac{1}{16 \pi G_{D}}\!\int\!\dd^{D}\!x
\sqrt{-g}\Big(-2\Lambda+\mathcal {L}_{1}+\alpha'_{2}\mathcal
{L}_{2}+\alpha'_{3}\mathcal {L}_{3}-4 \pi G_{D}
F_{\mu\nu}F^{\mu\nu}\Big),
\end{equation}
where
\begin{eqnarray}
&&\mathcal {L}_{1}=R,\nonumber\\
&&\mathcal{L}_{2}=R_{\mu\nu\gamma\delta}R^{\mu\nu\gamma\delta}-4R_{\mu\nu}R^{\mu\nu}+R^2,\nonumber\\
&&\mathcal{L}_{3}=2R^{\mu\nu\sigma\kappa}R_{\sigma\kappa\rho\tau}R^{\rho\tau}_{~\mu\nu}
+8R^{\mu\nu}_{~\sigma\rho}R^{\sigma\kappa}_{~\nu\tau}R^{\rho\tau}_{~\mu\kappa}
+24R^{\mu\nu\sigma\kappa}R_{\sigma\kappa\nu\rho}R^{\rho}_{~\mu}
+3RR^{\mu\nu\sigma\kappa}R_{\sigma\kappa\mu\nu}\nonumber\\
&&+24R^{\mu\nu\sigma\kappa}R_{~\sigma\mu}R_{~\kappa\nu}+16R^{\mu\nu}R_{\nu\sigma}R^{\sigma}_{~\mu}
-12RR^{\mu\nu}R_{\mu\nu}+R^3,
\end{eqnarray}
$\Lambda$ is the cosmological constant, $\alpha'_2$ and $\alpha'_3$
are Gauss-Bonnet and third order Lovelock coefficients,
respectively. The field strength is defined as
$F_{\mu\nu}(x)=\del_\mu A_\nu(x)-\del_\nu A_\mu (x)$. The
thermodynamics and geometric properties of black objects in Lovelock
gravity were studied in several papers
\cite{zanelli,caicao,deh2,gs}.
From the action (\ref{lovelockaction}), we can write down the
equation of motion\cite{mann},
\begin{eqnarray}
\label{einstein} \sum^{3}_{k=0}\frac{1}{2^{k+1}}c_{k}\delta^{\mu
c_1...c_k d_1...d_k}_{\nu e_1...e_k f_1...f_k}R^{e_1 f_1}_{~~c_1
d_1}\cdot\cdot\cdot R^{e_k f_k}_{c_k d_k}=8\pi G_{D} T^{\mu}_{~\nu},
\end{eqnarray}
where $ T^{\mu}_{~\nu} =F^{\mu}_{\rho}F_{\nu\sigma}g^{\rho\sigma}
-\frac{1}{4}g^{\mu}_{~\nu}F_{\rho\sigma}F^{\rho\sigma}$. Note that
for third order Lovelock gravity, we must deal with $D$-dimensional
spacetimes with $D\geq 7$.

If we choose
\begin{equation}
\label{alpha} \alpha'_2=\frac{\alpha}{(D-3)(D-4)},
~~\alpha'_3=\frac{\alpha^2}{3(D-3)\cdot\cdot\cdot(D-6)},
\end{equation}
the charged black hole solution in $D$ dimensions for this action is
described by~\cite{deh}
\begin{subequations}
\begin{eqnarray}
\label{metric} \dd s^2 &=& \displaystyle -H(r)N^2\dd
t^2+H^{-1}(r)\dd r^2+\frac{r^2}{l^2} h_{ij}\dd x^{i}\dd x^{j},
\\
A_t &=& \displaystyle -\frac{Q}{4\pi(D-3)r^{D-3}},
\end{eqnarray}
\end{subequations}

\vspace*{-6mm} \noindent with
$$
H(r)=k_{0}+\frac{r^2}{\alpha}\left\{1-\left[1-\frac{3\alpha}{l^2}\left(1-\frac{m
l^2 }{r^{D-1}}+\frac{q^2 l^2}{r^{2D-4}}\right)\right]^{1/3}
\right\}, \quad \Lambda =-\frac{(D-1)(D-2)}{2l^2},
$$
where  the parameter $l$ corresponds to AdS radius. The constant
$N^2$ will be fixed later. Note that the constant value of $k_0$ can
be $\pm1$ or $0$ and $h_{ij}\dd x^{i}\dd x^{j}$ represents the line
element of a $(D-2)$-dimensional hypersurface with constant
curvature $(D-2)(D-3)k_{0}$ and volume $V_{D-2}$. The gravitational
mass $M$ and the charge $Q$ are expressed as
\begin{eqnarray*}
M&=&\frac{(D-2)V_{D-2}}{16 \pi G_D  }m,
\\
Q^2&=&\frac{2\pi (D-2)(D-3)}{ G_D  }q^2.
\end{eqnarray*}
Taken the limit $\alpha'_2, \alpha'_3 \rightarrow 0$ with $k_0=0$,
the solution  corresponds to one for Reissner-Nordstr\"om-AdS
(RN-AdS). The hydrodynamic analysis in this background has been done
in\cite{gmsst1,msst}.

One may notice that here we use a black hole solution by choosing
particular values of $\alpha'_2$ and $\alpha'_3$ so that our
computation can be simplified greatly. Eq.(\ref{einstein})with the
choice (\ref{alpha}) yields one real and two complex solutions. We
use the real solution in (\ref{metric}). The general solution of
third order Lovelock gravity in $D$ dimensions for any arbitrary
values of $\alpha'_2$ and $\alpha'_3$ was obtained in \cite{deh},
but the line element of the metric turns out to be very complicated.
Furthermore, the general solution may present naked singularities,
which is not what we are interested \cite{deh}. In this paper, we
only focus on the special case given in (\ref{metric}).

 In the
following, we mainly focus on $D$-dimensional case with $k_0=0$.
Defining $\lambda=\alpha/l^2$, the function $H(r)$ becomes
\begin{equation}
\label{m} H(r)=\frac{r^2}{\lambda l^2}\left\{1-\left[1-3\lambda
\left(1-\frac{r^{D-1}_{+}}{r^{D-1}}-a\frac{r^{D-1}_{+}}{r^{D-1}}+a\frac{r^{2D-4}_{+}}{r^{2D-4}}\right)
\right]^{1/3}  \right\},
\end{equation}
where $a=\frac{q^2 l^2}{r^{2D-4}_{+}}$. The event horizon is located
at $r=r_{+}$. The constant $N^2$ in the metric (\ref{metric}) can be
fixed at the boundary whose geometry would reduce to flat Minkowski
metric conformaly, i.e.\ $\dd s^2\propto -c^2\dd t^2+\dd\vec{x}^2$.
On the boundary $r\rightarrow\infty$, we have
$$
H(r)N^2 \rightarrow\frac{r^2}{l^2},
$$
so that $N^2$ is found to be
\begin{equation}
N^2=\frac{\lambda}{1-(1-3\lambda)^{1/3}}.
\end{equation}
Note that the boundary speed of light is specified to be unity
$c=1$. From (\ref{m}), one can assume $\lambda\leq 1/3$ in order to
work with the branch of well-behaved solutions. That is because when
$\lambda> 1/3$ the sign of $H(r)$ becomes minus in the asymptotic
infinity, and we cannot recover the AdS geometry \footnote{In the
Gauss-Bonnet case, the function $H(r)$ has a different from:
$H(r)=\frac{r^2}{2\lambda l^2}\left\{1-\sqrt{1-4\lambda_{GB}
(1-\frac{r^2_{+}}{r^2})(1-\frac{r^2_{-}}{r^2})(1-\frac{r^2_{0}}{r^2})}
\right\}$, which implies that the significant value of $\lambda$
lies in the range $\lambda_{GB}\leq 1/4$. Beyond this point, the
Einstein-Maxwell-Gauss-Bonnet action does not admit a vacuum AdS
solution, and then the AdS/CFT correspondence is undefined. In
\cite{ges}, it was found that causality requires exactly
$\lambda_{GB}\leq 1/4$ in the $D\rightarrow \infty$ limit. This
result matches precisely the assumption (i.e. $\lambda_{GB}\leq
1/4$) used in \cite{shenker, fada}.}. In section 4, we will carry
out the causality analysis and  find the causality constraints
imposed on the value of $\lambda$.

We shall give thermodynamic quantities of this background. The
temperature at the event horizon is defined as
\begin{equation}
T=\frac{1}{2\pi\sqrt{g_{rr}}}\frac{\dd \sqrt{g_{tt}}}{\dd
r}=\frac{Nr_{+}}{4\pi l^2}\left[(D-1)-(D-3)a\right].
\end{equation}
The black brane approaches extremal when $a\rightarrow
\frac{D-1}{D-3}$ (i.e.\ $T\rightarrow 0$). The entropy of black
branes with $k_0=0$ obeys the area law \cite{deh} and thus the
entropy density has the form,
\begin{equation}
s=\frac{r^{D-2}_{+}}{4G_D l^{D-2}}.
\end{equation}

\section{Viscosity to entropy density ratio}
\setcounter{equation}{0} \setcounter{footnote}{0} We  explored the
charge dependence of $\eta/s$ in the presence of Gauss-Bonnet terms
for D-dimensional AdS black branes in \cite{ges}. In this section,
we generalize the previous result on $\eta/s$ \cite{gmsst,ges} to
 third order Lovelock gravity. It is convenient to introduce
coordinate in the following computation
\begin{eqnarray}
&&z=\frac{r}{r_{+}},
~~\omega=\frac{l^2}{r_{+}}\bar{\omega},~~k_{3}=\frac{l^2}{r^2_{+}}\bar{k}_3,~~
f(z)=\frac{l^2}{r^2_{+}}H(r), \nonumber\\
&&f(z)=\frac{z^2}{2
\lambda}\bigg[1-\left({1-3\lambda\bigg(1-\frac{a+1}{z^{D-1}}+\frac{a}{z^{2D-4}}}\bigg)\right)^{1/3}\bigg]
\end{eqnarray}
We now study the tensor type perturbation
$h^{x_1}_{x_2}(t,x_3,z)=\phi(t,x_3,z)$ on the black brane background
of the form
$$
ds^2=-f(z)N^2\dd t^2+\frac{\dd z^2}{f(z)}+\frac{z^2}{
l^2}\left(2\phi(t,x_3,z)\dd x_1 \dd x_2+\sum^{D-2}_{i=1}\dd
x^2_{i}\right),
$$
 Using Fourier decomposition
$$
\phi(t, x_3,z) = \!\int\!\frac{\dd^{D-1}k}{(2\pi)^{D-1}}
\mbox{e}^{-i\bar{\omega} t+i\bar{k}_{3}x_3}\phi(k, z),
$$
and expanding the action for tenor type gravitational perturbations
$\phi(t,x_3,z)$ to the second order, we obtain the effective action
in the momentum space
\begin{equation}\label{action}
S=\frac{1}{16 \pi G_D}\int \frac{d\omega d k_3}{(2\pi)^{D-3}}dz
\sqrt{-g}\left(M(z)\phi'\phi'+M_{2}(z){\phi}^2\right),
\end{equation} where the prime denotes the derivative with respect to
$z$.

An easy way to obtain the equation of motion of the tensor type
perturbation is to substitute the fluctuated metric into Eq.
(\ref{einstein}). One then find
  the linearized equation of motion
for $\phi(z)$ from the third order Lovelock field equation:
\begin{equation}\label{maineq}
M(z) \phi''(z)+M'(z)\phi'(z)+M_2 \phi(z)=0
\end{equation}
where
\begin{eqnarray}
&&M(z)=z^{D-2}f\left\{1-\frac{2 \lambda}{D-3}
\left[z^{-1}f'+z^{-2}(D-5)f\right]+\frac{\lambda^2
z^{-3}}{D-5}\left[2f'+(D-7)z^{-1}f'\right]f\right\}\nonumber
\\
&&M_2=M(z)\frac{\omega^2}{N^2f^2}-k^2_{3}z^{D-4}\times\nonumber
\\
&&\left\{1-\frac{2\lambda}{(D-3)(D-4)}\left(f''+(D-5)(D-6)z^{-2}f+2(D-5)z^{-1}f'\right)
\right. \nonumber\\
&&\left.+\frac{2\lambda^2}{(D-3)(D-4)}\left[z^{-3}(f'^2+ff'')+2(D-7)z^{-4}ff'+\frac{1}{2}(D-7)(D-8)z^{-5}f^2\right]
\right\},
\end{eqnarray}
We would like to emphasize that when $D=5$ and the $\lambda^2$ terms
vanished, (\ref{maineq}) reduces to the main equation obtained in
\cite{shenker,gmsst}. The shear viscosity involves physics in the
lower frequency and lower momentum limit and one can neglect the
$M_2 (z)$ term in solving Eq.(\ref{maineq}). For the convenient
calculation of the shear viscosity, we would like to introduce a new
variable $u=\frac{1}{z}$ and rewrite equation (\ref{maineq}) in the
new coordinate
\begin{equation}\label{newmaineq}
J(u) \phi''(u)+J'(u)\phi'(u)+J_2(u) \phi(u)=0,
\end{equation}
where $J(u)=\frac{M(1/z)}{z^2}$ and $J_2(u)=M_2(\frac{1}{z})$.

In order to solve the equation of motion (\ref{newmaineq}) in
hydrodynamic regime, let us assume that the solution yields
\begin{equation}
\label{first}
\phi\left(u\right)=\left(1-u\right)^{\nu}F\left(u\right),
\end{equation}where $F\left(u\right)$ is  regular at the
horizon. $\nu=\pm i \frac{\omega}{4 \pi T}$ can be fixed by
substituting (\ref{first}) into the equation of motion, which we
choose
$$
\nu=-i \frac{\omega}{4 \pi T},
$$
Since we only need to know the behavior at $\omega\rightarrow 0$
region, it is sufficient to expand the solution in terms of
frequencies up to the linear order of $\nu$
\begin{equation}
F(u) =F_{0}(u)+\nu F_{1}(u)+ {\cal O}(\nu^2, k^2_3).
\end{equation}
The equation governing $F_{0}(u)$ goes as
\begin{equation}
\left[J(u)F'_0(u)\right]'=0,
\end{equation}
and can be solved as
\begin{equation}
F'_0(u)=\frac{C_1}{J(u)},
\end{equation}
where $C_1$ is an integration constant and must be zero as $J(u)$
goes zero at the horizon so that $F_0(u)$ is regular at the horizon.
Therefore, $F_0(u)$ is a constant, i.e. $F_0(u)=C$. From the
equation at $\mathcal {O}(\nu)$,
\begin{equation}
\left[J(u)F'_1(u)\right]'-\left(\frac{C}{1-u}J(u)\right)'=0,
\end{equation}
we find that the solution can be written as
\begin{equation}
F'_1(u)=\frac{C}{1-u}+\frac{C_2}{J(u)}.
\end{equation}
Regularity of $F_1(u)$ at the horizon requires that
\begin{equation}
C_2=-\bigg[\left((D-1)-(D-3)a\right)(1-\frac{2\lambda}{D-3}
((D-1)-(D-3)a))\bigg]C.
\end{equation}
The value of $C$ can fixed by the boundary condition
$C=\lim_{u\rightarrow 0}\phi(u)=1$. It is worth to noting that the
above calculation is same as the Gauss-Bonnet cases given in
\cite{gmsst,ges}.

Using the equation of motion, we write down the on-shell action
\begin{equation}
I_{on-shell}=-\frac{r^{D-1}_{+}N}{16\pi G_D l^D}
\!\int\!\frac{\dd^{D-1} k}{(2\pi)^{D-1}}
\Big(J(u)\phi(u)\phi'(u)+\cdots\Big)\Bigg|_{u=0}^{u=1}..
\end{equation}
The shear viscosity can be read off using the Kubo formula
\begin{eqnarray}
\eta &=&\lim_{\omega\rightarrow 0}\frac{\rm Im
G(\omega,0)}{\omega}=\frac{r^{D-1}_{+}N}{16\pi G_D
l^D}\lim_{\omega\rightarrow 0}\frac{J(u)\phi(u)\phi'(u)|_{u=0}}{i
\omega}\nonumber\\ &=&\frac{1}{16 \pi
G_D}\left(\frac{r^{D-2}_{+}}{l^{D-2}}\right)\Big(1-\frac{2\lambda}{D-3}
[(D-1)-(D-3)a]\Big).
\end{eqnarray}

 The ratio of the shear viscosity to the entropy density turns
out to be
\begin{equation}
\frac{\eta}{s} =\frac{1}{4 \pi } \left(1-\frac{2\lambda}{D-3}
[(D-1)-(D-3)a]\right).
\end{equation}
We obtain the same result as that of \cite{ges}, which is also
consistent with the prediction made in \cite{brustein1} when $a=0$.
In other words, the third order Lovelock coupling constant
$\alpha^{'}_3$ (or $\lambda^2$ in our case) does not contribute
 to the shear viscosity.
\section{Causality }
\setcounter{equation}{0} \setcounter{footnote}{0} The shear
viscosity above is calculated in the hydrodynamical regime ( i.e.
$k^{\mu}\rightarrow 0$). In this and next sections, we will turn to
the high frequency regime ( i.e. $k^{\mu}\rightarrow \infty$) and
discuss the causality issue.

Due to higher derivative terms in the gravity action, the equation
(\ref{maineq}) for the propagation of a transverse graviton differs
from that of a minimally coupled massless scalar field propagating
in the same background geometry. Writing the wave function as
\begin{equation}
\label{phi} \phi(x,u)=\mbox{e}^{-i\omega t+ikz+ik_{3}x_3},
\end{equation}
and taking large momenta limit $k^\mu\rightarrow\infty$, one can
find that the equation of motion (\ref{maineq}) reduces to
\begin{equation}
\label{effeq} k^{\mu}k^{\nu}g_{\mu\nu}^{\rm eff}\simeq 0,
\end{equation}
where the effective metric is given by
\begin{equation}
\dd s^2_{\rm eff} =g^{\rm eff}_{\mu\nu}\dd x^\mu\dd x^\nu ={N^2
f(z)} \left(-\dd t^2+\frac{1}{c^2_g}\dd x^2_3\right)
+\frac{1}{f(z)}\dd z^2.
\end{equation}
Note that $c^2_g$ can be interpreted as the local speed of graviton:
\begin{equation}
\label{cg2} c^2_g(z)=\frac{N^2 f}{z^2} \frac{ h_{1}}{h_{2}},
\end{equation}
where
\begin{eqnarray}
h_{1}&=&\left\{1-\frac{2\lambda}{(D-3)(D-4)}\left(f''+(D-5)(D-6)z^{-2}f+2(D-5)z^{-1}f'\right)
\right. \nonumber\\
&&\left.+\frac{2\lambda^2}{(D-3)(D-4)}\left[z^{-3}(f'^2+ff'')+2(D-7)z^{-4}ff'+\frac{1}{2}(D-7)(D-8)z^{-5}f^2\right]
\right\},\nonumber\\
h_{2}&=&\left\{1-\frac{2 \lambda}{D-3}
\left[z^{-1}f'+z^{-2}(D-5)f\right]+\frac{\lambda^2
z^{-3}}{D-5}\left[2f'+(D-7)z^{-1}f'\right]f\right\}.\nonumber
\end{eqnarray}
The above equations can exactly reduce to Gauss-Bonnet cases found
in \cite{shenker, shenker1}, if the  $\lambda^2$ terms vanished.
 For $D=10$, we can expand $c^2_g$
near the boundary $\frac{1}{z}=0$,
\begin{eqnarray}
\label{exp} c^2_g-1=
-\frac{4\left[1-(1-3\lambda)^{1/3}\right]^2}{4(1-3\lambda)^{2/3}+2(1-3\lambda)^{1/3}-1}+\mathcal
{O}(\frac{1}{z}).
\end{eqnarray}
 We can see that $c^2_g-1$ is always negative, which
implies that the local speed of graviton is smaller than the local
speed of light of the boundary CFT without any charge dependence .
One can further check that for $D=7,8,9...$, the first term  in
$c^2_g-1$ is also negative. Therefore, from (\ref{exp}) we can see
that the causality imposes no constraints on possible values of
$\lambda$. Figure 1 demonstrates that the value of $c^2_g$ lies in
the region $0.80\leq c^2_g \leq -0.8$ as a function of $u$ and
$\lambda$. Following the procedure of Ref.\cite{shenker1}, one can
find that the group velocity of the graviton is given by
\begin{equation}
v_g=\frac{\dd\omega}{\dd k}\sim c_g.
\end{equation}
So different from the Gauss-Bonnet cases \cite{shenker1,gmsst},
there is no causality violation in third order Lovelock theory with
the particular choice of $\alpha^{'}_{2}$ and $\alpha^{'}_{3}$ . The
difference comes from the fact that $\alpha^{'}_{3}$ terms change
the causal structure of the boundary $CFT$. For more than third
order Lovelock theory, the causal structure might be further
modified
 by $\alpha'_i (i>3)$.

 \begin{figure}[htbp]
\begin{minipage}{1\hsize}
\begin{center}
\includegraphics*[scale=0.53] {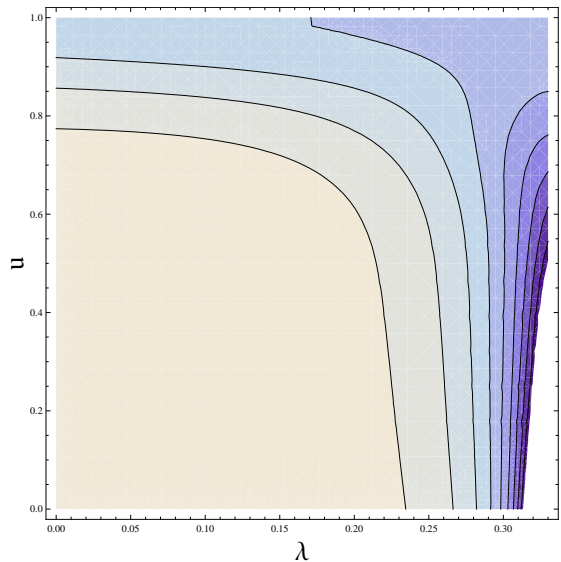}
\end{center}
\caption{ $c^2_g$ as a function of $u$ and $\lambda$ when we choose
$D=7$ and $a=1.4$. The lines correspond to $0.8, 0.6,...,-0.8$,
respectively, from left to right.} \label{1}
\end{minipage}
\end{figure}
One may notice that our discussions on the causal structure of
 third order Lovelock gravity are based on the metric (\ref{metric})
 and the equation motion in high frequency limit (\ref{effeq}).
 Hence, if we consider
 the general solution of
third order Lovelock gravity with arbitrary values of
$\alpha^{'}_{2}$ and $\alpha^{'}_{3}$, we may find totally different
causal structure. It remains to be carried out in the future.

\section{Instability}
\setcounter{equation}{0} \setcounter{footnote}{0} In section 4, we
have demonstrated that for the RN-AdS black brane in third order
Lovelock theory, causality violation does not happen, which implies
that the results obtained in \cite{shenker1,gmsst,ges} might not be
so universal as we expected. In this section, we extend our previous
work on black brane stability to third order Lovelock gravity.

 Now, we rewrite the main equation in a Schr$\rm\ddot{o}$dinger form,
\begin{equation}
-\frac{d^2 \psi}{dr^2_{*}}+V\left(z(r_{*})\right)\psi=\omega^2 \psi,
~~~\frac{dr_{*}}{dz}=\frac{1}{Nf(z)},\label{schr}
\end{equation}
where $\psi\left(z(r_{*})\right)$ and the potential is defined by
\begin{eqnarray}
&&\psi =K(z)\phi,~~~K(z)\equiv\sqrt{\frac{M(z)}{N f(z)}},
V=k^2c^2_g+V_{1}(z),\nonumber\\ &&V_{1}(z)\equiv
N^2\left[\left(f(z)\frac{\partial \ln K(z)}{\partial
z}\right)^2+f(z)\frac{\partial}{\partial z}\left(f(z)\frac{\partial
\ln K(z)}{\partial z}\right)\right]
\end{eqnarray}
In the large momentum limit, the Schr$\rm \ddot{o}$dinger potential
develops a negative gap near the horizon and the negative-valued
potential in turn leads to instability of the black brane. In the
large momenta limit $k^{\mu}\rightarrow \infty$, the potential is
mainly contributed by $k^2 c^2_g$. For charged black branes, $c^2_g$
can be negative near the horizon and the potential is deep enough to
have bound states living there. The negative-valued potential yields
negative energy eigenvalue (i.e. $\omega^2<0$).  The imaginary part
of $\omega$ can then be positive. Substituting the eigenvalue of
$\omega$ to the wave function for tensor type perturbations, one
immediately find that  perturbations grow as time goes on and the
black branes thus are unstable. The negative-valued energy bound
state corresponds to modes of tachyonic mass Minkowski slices and
signals an instability of the black brane \cite{troost}.
 Let us expand $c^2_g$ in series
of $(1-u)$,
\begin{eqnarray}
c^2_g=&& N^2 \left[(D-1)-(D-3)a\right]\left\{D^2
\left[2\lambda^2(a-1)^2+2(a+1)\lambda-1\right]\right.\nonumber\\
&&\left.-D\left[4\lambda^2(3a^2-4a+1)+2\lambda(a+7)-7\right]+\left[\lambda^2(3a-1)^2-6\lambda(a-1)-6\right]
\right\}\nonumber\\
&&\left\{(D-4) \left[(D-3)-2\lambda \left(D-1-(D-3)a\right)
\right]\right\}^{-1}(1-u)+\mathcal{O}((1-u)^2).
\end{eqnarray}
Since $0\leq a\leq \frac{D-3}{D-1}$, and $0\leq u \leq 1$, $c^2_g$
will be negative, if
\begin{eqnarray}
&&\left\{D^2
\left[2\lambda^2(a-1)^2+2(a+1)\lambda-1\right]\right.\nonumber\\
&&\left.-D\left[4\lambda^2(3a^2-4a+1)+2\lambda(a+7)-7\right]+\left[\lambda^2(3a-1)^2-6\lambda(a-1)-6\right]
\right\}\nonumber\\
&&\left\{(D-4) \left[(D-3)-2\lambda \left(D-1-(D-3)a\right)
\right]\right\}^{-1} <0.
\end{eqnarray}
From the above formula, we find the critical value of $\lambda$,
\begin{eqnarray}
\label{stability}
\lambda_{\rm c}&& =\frac{1}{2}\bigg\{-(D-1)(D-6)-(D-3)(D+2)a \nonumber\\
&&+\bigg\{(D-1)^2(3D^2-26D+60)+(D-3)^2(3D^2-10D+28)a^2\nonumber\\
&&-2a(D^4-14 D^3+79 D^2-174
D+108)\bigg\}^{\frac{1}{2}}\bigg\}\bigg\{D-1-(D-3)a\bigg\}^{-2} .
\end{eqnarray}
Above the line of $\lambda_{c}$, $c^2_g$ can be negative.
Eq.(\ref{stability}) tells us that the stability of the black brane
depends on the charge. The minimal value of $\lambda_{\rm c}$ can be
obtained in the limit $a\rightarrow (\frac{D-1}{D-3})$,
\begin{equation}
\lambda_{\rm c, \ min} = \frac{1}{4}\frac{(D-3)(D-4)}{(D-1)(D-2)}.
\end{equation}
When $D=5$, we recover the result found in \cite{gmsst}. Usually,
for the application of AdS/CFT correspondence, we do not need to
take infinite dimensionality limit. But the stability of higher
dimensional black holes itself is an important topic in the study of
black hole physics.

 The Einstein-Hilbert action is just the first term in
the derivative expansion in a low energy effective theory. The
Gauss-Bonnet and the third order Lovelock terms can be regarded as
higher order corrections to the Einstein gravity. In this sense, the
higher derivative gravity coupling constants should be small. In our
discussions, we have found that the coupling constant $\lambda$
($=(D-3)(D-4)\alpha/l^2$) depends on the dimensionality $D$. But it
seems that for fixed $\alpha$ and AdS radius $l$ as $D$ approaches
infinity, $\lambda$ would be very large. That is not what we want.
By doing stability analysis, we will find a way to restrict the
value of $\lambda$. As the value of $D$ increases, one finds that
$\lambda_{\rm c, \ min}$ is bounded by $1/4$,
\begin{equation}
\lim_{(D,a)\rightarrow(\infty,\frac{D-1}{D-3})}\lambda_{\rm
c}=\frac{1}{4}
\end{equation}
Thus we reproduce the result of \cite{ges}.  Third order Lovelock
gravity in our case does not add new constraints on the stability of
the black brane.
 \begin{table*}[htbp]
\begin{center}
\begin{tabular}{|c|c|c|c|c|c|c|c|c|}
\hline
$D$&$\lambda=0.33$&$\lambda=0.30$&$\lambda=0.26$&$\lambda=0.22$&$\lambda=0.18$\\
\hline
$7$&$ 35.7830 \imo$&$34.2779\imo$&$ 28.5647\imo$&$21.3057\imo $&$10.0058 \imo$\\
$8$&$ 27.8612 \imo$&$ 31.3230\imo$&$22.1646$&$ 14.6227 \imo$&$5.1492 \imo$\\
$9$&$ 23.4121\imo$&$ 22.7276\imo$&$17.9269 \imo$&$ 11.2243 \imo$&$3.1228 \imo$\\
$10$&$ 22.0265\imo$&$ 18.5886\imo$&$12.2811 \imo$&$ \rm -$&$- $\\
\hline
\end{tabular}
\caption{Unstable QNMs for third order charged Lovelock black brane
perturbation of tensor type for fixed charge ($a=1.20$) and
$k_3=500$. As $D$ increases, the unstable modes are suppressed. And
also, small $\lambda$ helps to smooth the perturbation. }\label{one}
\end{center}
\end{table*}
\begin{table*}[htbp]
\begin{center}
\begin{tabular}{|c|c|c|c|c|c|c|c|c|}
\hline
$\lambda$&$a=1.3$&$a=1.2$&$a=1.0$&$a=0.8$&$a=0.6$\\
\hline
$0.33$&$  31.6732 \imo$&$  27.8612\imo$&$12.7312 \imo$&$  \rm -$&$\rm -$\\
$0.30$&$ 31.5500 \imo$&$27.3852\imo$&$ 9.9506\imo$&$ \rm -$&$ \rm -$\\
$0.27$&$  28.1982\imo$&$ 23.7183\imo$&$ 5.1296 \imo $&$ \rm -$&$ -$\\
$0.24$&$ 23.4156\imo$&$ 18.6473 \imo$&$\rm -$&$ \rm -$&$ -$\\
\hline
\end{tabular}
\caption{Unstable QNMs for third order charged Lovelock black brane
perturbation of tensor type for fixed dimensionality ($D=8$) and
$k_3=500$. This table indicates that instability is increased by a
chemical potential.}\label{two}
\end{center}
\end{table*}
\begin{table*}[htbp]
\begin{center}
\begin{tabular}{|c|c|c|c|c|c|c|c|c|}
\hline
$D$&$a=1.4$&$a=1.3$&$a=1.2$&$a=0.8$\\
\hline
$7$&$ 42.7952\imo$&$39.3407 \imo$&$  33.3140\imo$&$\rm -$\\
$8$&$ 30.8934 \imo$&$29.5030\imo$&$ 25.1211\imo$&$ \rm -$\\
$9$&$  \rm -$&$ 23.4155\imo$&$ 20.6184\imo$&$ \rm -$\\
$10$&$ \rm -$&$ \rm -$&$17.442\imo$&$ \rm -$\\
\hline
\end{tabular}
\caption{Unstable QNMs for third order charged Lovelock  black brane
perturbation of tensor type for fixed $\lambda$ ($\lambda=0.28$).
This table shows combined effects of D and the chemical potential.
Note that $a=1.4$ exceeds the maximal value of charge permitted for
$9$- and $10$-dimensional charged black brane and thus we leave the
frequency blank there.}\label{three}
\end{center}
\end{table*}
To show explicitly the behavior of gravitational perturbation in
higher dimensions ($D\geq 7$), we solve the Schr$\rm \ddot{o}$dinger
equation (\ref{schr}) with negative valued potential numerically and
find unstable quasinormal modes (QNMs)(see tables \ref{one},
\ref{two} and \ref{three} ).

Table \ref{one} demonstrates that the unstable modes of the black
branes are suppressed as $D$ increases. Table 2 and 3 tell us the
same story as we found in \cite{ges}, that is to say, lower value of
charge-($a$) and $\lambda$ stabilize the perturbation, while the
lower value of $D$ strengthens the instability. The reason for why
higher $D$ suppresses the gravitational fluctuation is because that
no matter how big $D$ is, $\lambda$  is bounded by $1/4$, which
means that for fixed AdS radius $l$, $\alpha^{'}\rightarrow 0$ as
the value of $D$ goes up. The upper bound of $\lambda$ constrains
the gravitational perturbation in the larger $D$ limit. For QNMs of
RN-AdS black holes in Einstein and Gauss-Bonnet gravity, one may
refer to \cite{wang,konoplya}.

It would be very interesting to check for fixed value of charge, for
which value of $\lambda$ the black brane becomes stable. In order to
do this, one should first fix $D$ in (\ref{stability}), then obtain
a formula between $\lambda$ and $a$. Actually, (\ref{stability})
indicates that for $\lambda< \lambda_{c}(D,a)$, the black brane
becomes stable. For $5$-dimensional black brane with charge in
Gauss-Bonnet gravity, constraints from causality as well as
stability separate the physics into four regions in $(a,\lambda)$
space: consistent region; only causality violation region; only
unstable modes region; both causality violation and unstable modes
region (see figure 4 in \cite{gmsst} for details). But for the
particular case we are considering here, since causality violation
does not occur, we have only two phases in the $(a,\lambda)$ space:
stable  and unstable modes regions marked by (\ref{stability}). One
thing one need to be aware of is that instability of the black brane
does not correspond to any fundamental pathology with the theory.
This is quiet different from the causality violation which means
that a theory is pathological. In the dual gravitational
description, the unstable QNMs is identified with unstable uniform
plasma with respect to certain non-uniform perturbation
\cite{Buchel}.

\section{Conclusions and discussions}
\setcounter{equation}{0} \setcounter{footnote}{0} In conclusion, we
derive  the main equation for tensor type perturbation in third
order Lovelock theory and compute  the shear viscosity. The result
turns out to be in agreement with the prediction made in
\cite{brustein1} when $a=0$, that is to say, the third order
Lovelock term does not add new ingredients into the shear viscosity
of Gauss-Bonnet theory.

We notice that an interesting point comes from the causality
analysis. While in the Gauss-Bonnet theory, causality could be
violated in the boundary $\rm CFT$, we do not find causality
violation in third order Lovelock theory. From (\ref{cg2}), we can
see that the local speed of graviton depends on both  $\alpha^{'}_2$
($\sim \lambda$) and $\alpha^{'}_3$ ($\sim \lambda^2$). Although we
are working only with a special choice of $\alpha^{'}_2$ and
$\alpha^{'}_3$, Eq. (\ref{cg2}) implies that causality receives
corrections from the $\alpha^{'}_3$  term. Thus, the causal
structure in general third order Lovelock gravity must be different
from the Gauss-Bonnet gravity. We also expect that higher than
fourth order Lovelock theory may  impose more constraints on the
causal structure of the boundary $\rm CFT$.

The instability of charged black brane with third order Lovelock
theory shows the same properties as that of Gauss-Bonnet
corrections. We find that higher $D$ suppresses the unstable modes,
but larger value of charge and $\lambda$ strengthen  the
perturbation. As $D$ approaches infinity, the stability requires
$\lambda$ to be bounded by $1/4$. This is an important observation
in that Eq.(\ref{m}) indicates that $\lambda$ could be as big as
$1/3$ without any causality violation happens in third order
Lovelock gravity. But $\lambda \sim 1/3$ violates the assumption
$\lambda_{GB}\leq 1/4$ used in \cite{shenker,fada}. Fortunately,
after imposing the stability constraint, we can recover the
requirement $\lambda\leq 1/4$ and thus  third order Lovelock and
Gauss-Bonnet gravity are somehow consistent.

\vspace*{10mm} \noindent
 {\large{\bf Acknowledgments}}

\vspace{1mm}This work is supported partly by the Shanghai Leading
Academic Discipline Project (project number S30105). The work of SFW
is partly supported by NSFC under Grant Nos. 10847102, and the
Innovation Foundation of Shanghai University.  The work of SJS was
supported by KOSEF Grant R01-2007-000-10214-0. This work is also
supported by Korea Research Foundation Grant KRF-2007-314-C00052 and
SRC Program of the KOSEF through the CQUeST with grant number
R11-2005-021. 

\vspace{1mm}


\begin{thebibliography}{99}
\bibitem{ads/cft}
J. M. Maldacena, {Adv. Theor. Math. Phys.} {\bf 2} (1998) 231, {\tt
[arXiv:hep-th/9711200]}.
\bibitem{gkp}
S. S. Gubser, I.R. Klebanov and A.M. Polyakov, Phys.\ Lett.\ {\bf
B428} (1998) 105, {\tt [arXiv:hep-th/9802109]}.
\bibitem{w}
E. Witten, Adv.\ Theor.\ Math.\ Phys.\ {\bf 2} (1998) 253, {\tt
[arXiv:hep-th/9802150]}.
\bibitem{pss0}
G. Policastro, D. T. Son and A.O. Starinets, Phys.\ Rev.\ Lett.\
{\bf 87} (2001) 081601, {\tt [arXiv:hep-th/0104066]}.
\bibitem{kss}
P. Kovtun, D. T. Son and A.O. Starinets,
JHEP {\bf 0310} (2003) 064, \\
{\tt [arXiv:hep-th/0309213]}.
\bibitem{bl}
A. Buchel and J. T. Liu,
Phys.\ Rev.\ Lett.\  {\bf 93} (2004) 090602, \\
{\tt [arXiv:hep-th/0311175]}.
\bibitem{ssz}
E. Shuryak, S.-J. Sin and I. Zahed,
J.\ Korean Phys.\ Soc.\  {\bf 50} (2007) 384, \\
{\tt [arXiv:hep-th/0511199]}.
\bibitem{ksz}
K.-Y. Kim, S.-J. Sin and I. Zahed, {\tt [arXiv:hep-th/0608046]}.
\bibitem{ht}
N. Horigome and Y. Tanii, JHEP {\bf 0701} (2007) 072, {\tt
[arXiv:hep-th/0608198]}.
\bibitem{nssy1}
S. Nakamura, Y. Seo, S.-J. Sin and K. P. Yogendran, {\tt
[arXiv:hep-th/0611021]}.
\bibitem{kmmmt}
S. Kobayashi, D. Mateos, S. Matsuura, R.C. Myers and R.M. Thomson,
JHEP {\bf 0702} (2007) 016, {\tt [arXiv:hep-th/0611099]}.
\bibitem{nssy2}
S. Nakamura, Y. Seo, S.-J. Sin and K. P. Yogendran, {\tt
[arXiv:0708.2818[hep-th]]}.
\bibitem{huang}J. W. Chen, M. Huang, Y.H. Li, E. Nakano and D. L.
Yang,
 Phys. Lett. B {\bf 670} (2008) 18, {\tt [arXiv:0709.3434 [hep-ph]]}
\bibitem{kovtun}
P. Kovtun, D. T. Son and A.O. Starinets, Phys.\ Rev.\ Lett.\  {\bf
94} (2005) 111601, {\tt [arXiv:hep-th/0405231]}.
\bibitem{kp}
Y. Kats and P. Petrov,  JHEP {\bf 0901} (2009) 044 {\tt
[arXiv:0712.0743[hep-th]]}.
\bibitem{shenker}
M. Brigante, H. Liu, R.C. Myers, S. Shenker and S. Yaida, Phys. Rev.
{\bf D77} (2008) 126006, {\tt [arXiv:0712.0805[hep-th]]}.
\bibitem{shenker1}
M. Brigante, H. Liu, R.C. Myers, S. Shenker and S. Yaida, Phys. Rev.
Lett. {\bf 100} (2008) 191601, {\tt [arXiv:0802.3318[hep-th]]}.
\bibitem{neupane}
I.P. Neupane and N. Dahhich, Class. Quant. Grav. 26 (2009)  015013.
{\tt [arXiv:0808.1919[hep-th]]}; I.P. Neupane, {\tt
[arXiv:0904.4805[hep-th]]}.
\bibitem{brustein1}R. Brustein and A. M. Medved, {\tt [arXiv:0808.3498[hep-th]]}
\bibitem{brustein2}R. Brustein and A. M. Medved, {\tt [arXiv:0810.2193[hep-th]]}
\bibitem{Iq} N. Iqbal and H. Liu, {\tt [arXiv:0809.3808[hep-th]]}
\bibitem{ban} N. Banerjee and S. Dutta, {\tt [arXiv:0903.3925[hep-th]]}
\bibitem{zw}B, Zwiebach, Phys. Lett. B {\bf 156} (1986) 315
\bibitem{zu}B, Zumino, Phys. Rep. {\bf 137} (1986) 109
\bibitem{lovelock} D. Lovelock, J. Math. Phys. {\bf 12} (1971) 498
\bibitem{gmsst}
X.~H. Ge, Y. Matsuo, F.-W. Shu, S.-J. Sin and T. Tsukioka, JHEP 0810
(2008) 009, {\tt [arXiv:0808.2354[hep-th]]}
\bibitem{ges} X.~H. Ge and S.-J. Sin, JHEP {\bf 0905}  (2009) 051{\tt [arXiv:0903.2527[hep-th]]}
\bibitem{cai2}R. G. Cai and Y. W. Sun, JHEP {\bf 0603} (2008) 052, {\tt [arXiv:0807.2377[hep-th]]}; R. G. Cai, Z. Y. Nie and Y. W. Sun, Phys. Rev. {\bf D78} (2008)126007  {\tt [arXiv:0811.1665[hep-th]]}
\bibitem{cai3}R. G. Cai, N. Ohta, Z. Y. Nie and Y. W. Sun, {\tt [arXiv:0901.1421[hep-th]]}
\bibitem{Jliu}S. Cremonini, K. Hanaki, J. T. Liu and P. Szepietowski, {\tt [arXiv:0903.3244 [hep-th]]}
\bibitem{merys}R. C. Myers, M. F. Paulos and A. Sinha, {\tt [arXiv:0903.2834  [hep-th]]}
\bibitem{fada}K. Bitaghsir Fadafan,
  JHEP 0812 (2008) 051, {\tt [arXiv:0803.2777 [hep-th]]};
  {\tt [arXiv: 0809.1336[hep-th]]}
\bibitem{McInnes} B. McInnes, {\tt [arXiv:0905.1180 [hep-th]]}
\bibitem{Buchel} A. Buchel and R. Myers, {\tt [arXiv:0906.2922 [hep-th]]}
\bibitem{dotti}
G. Dotti and R.J. Gleiser, Phys.\ Rev.\ {\bf D72} (2005) 044018,
{\tt [arXiv:gr-qc/0503117]}; \\
R.J. Gleiser and G. Dotti, Phys.\ Rev.\  {\bf D72} (2005) 124002,
{\tt [arXiv:gr-qc/0510069]}; \\
M. Beroiz, G. Dotti and  R.J. Gleiser,
Phys. Rev. {\bf D76} (2007) 024012, \\
{\tt [arXiv:hep-th/0703074]}.
\bibitem{konoplya}
R.A. Konoplya and A. Zhidenko, Phys. Rev. {\bf D77} (2008) 104004,
{\tt [arXiv:0802.0267]}.
\bibitem{taka} T. Takahashi and J. Soda,  Phys. Rev. {\bf D75}
(2009) 104025, {\tt arXiv:0902.2921 [gr-qc]}; {\tt arXiv:0907.0556
[gr-qc]}
\bibitem{deh} M. H. Dehghani and M. Shamirzaie, Phys. Rev. {\bf D
72} (2005) 124015
\bibitem{zanelli} J. Crisostomo, R.
Troncoso and J. Zanelli, Phys. Rev. D {\bf 62} (2000) 084013, {\tt
[arXiv:hep-th/0003271]}; R. Aros, R. Troncoso and J. Zanelli, Phys.
Rev. D {\bf 63} (2001) 084015, {\tt [arXiv:hep-th/0011097]}.
\bibitem{caicao} R. G. Cai, Phys. Lett. B {\bf 582} (2004) 237, {\tt
[arXiv:hep-th/0311240]}; R. G. Cai, N. Ohta, Phys.Rev. D74 (2006)
064001,  {\tt [arXiv:hep-th/0604088]}; R. G. Cai, L. M. Cao, Y. P.
Hu and S. P. Kim, Phys.Rev.D {\bf 78} (2008)124012.
\bibitem{deh2} M. H. Dehghani and N. Farhangkhah,  Phys. Rev. D {\bf 78} (2008)
064015; M. H. Dehghani, N. Bostani and S. H. Hendi, Phys. Rev. D
{\bf 78} (2008) 064031
\bibitem{gs}
F.-W. Shu and X. H. Ge, JHEP {\bf 0808} (2008) 021 {\tt
[arXiv:0804.2724 [hep-th]]}; {\tt [arXiv:0804.2123 [hep-th]]}.
\bibitem{mann}D. Kastor and R. Mann, {\tt [arXiv:hep-th/0603168]}
\bibitem{gmsst1}
X. H. Ge, Y. Matsuo, F.-W. Shu, S.-J. Sin and T. Tsukioka, Prog.
Theor. Phys. 120 (2008) 833, [arXiv:0806.4460[hep-th]].
\bibitem{msst}
Y. Matsuo,  S.-J. Sin, S. Takeuchi, T. Tsukioka and C. -M. Yoo, {\tt
[arXiv:0901.0610 [hep-th]]}.
\bibitem{troost} J. Troost, Phys. Lett. {\bf B 578} (2004) 210 , [hep-th/0308044].
\bibitem{wang} B. Wang, C. Y. Li and E. Abdalla, Phys. Lett. B {\bf
481} (2000) 79
\end{thebibliography}
\end{document}